\documentclass[reprint,noshowpacs,noshowkeys,prd,balancelastpage,nofootinbib]{revtex4}
\usepackage{amsmath,amsfonts,amssymb}
\usepackage{latexsym}
\usepackage{dcolumn}
\usepackage{graphicx,epsfig}
\usepackage{amsthm}
\usepackage{color}
\usepackage{graphicx}
\usepackage[unicode=true,
 bookmarks=false,
 breaklinks=false,pdfborder={0 0 1},backref=section,colorlinks=true]
 {hyperref}
\hypersetup{
 linkcolor=blue,urlcolor=blue,citecolor=blue}
\begin{document}
\title{IR-deformed thermodynamics of quantum bouncers and the issue of dimensional reduction}
\author{P. Dehghani}
\email{parham.dehghani@emu.edu.tr (Corresponding Author)}

\affiliation{Department of Physics, Faculty of Arts and Sciences, Eastern Mediterranean
University, Famagusta, Northern Cyprus via Mersin 10, Turkey~~\\
 }

\author{K. Nozari}
\email{knozari@umz.ac.ir}

\affiliation{Department of Physics, Faculty of Basic Sciences,\\
University of Mazandaran,\\
P. O. Box 47416-95447, Babolsar, Iran\\
 }
\begin{abstract}
We probe the low-temperature behavior of a system of quantum bouncers as a theoretical model for ultracold neutrons within a low energy modified version of the standard quantum mechanics, due to the gravitational effects. Working in one dimension, the energy spectrum and bound states of a deformed quantum bouncer are obtained using the first-order WKB approximation, granted the very low energy regime of the particle. In this manner, we can study energy levels of a system of ultracold neutrons as an informative probe towards exploring the low energy manifestation of semi-classical quantum gravitational effects. Our calculated energy levels of ultracold neutrons are in accordance with the observed energy levels, as obtained in the famous Nesvizhevsky \emph{et al.} experiment, with a negative constant deformation, as dependent on the deformation parameter. In advance, we tackle modified thermodynamics of a system of quantum bouncers in the infrared regime via an ensemble theory both in one dimension and also three dimensions, to seek for any trace of an effective, thermodynamic dimensional reduction in this low energy regime of semi-classical quantum gravity. While the issue of dimensional reduction has been essentially assigned to the high energy regime, here we show that there is a trace of an effective, thermodynamic dimensional reduction in infrared regime with one important difference: in the high energy regime, the dimensional reduction effectively occurs from $D=3$ to $D=1$, but here, in this low energy regime, there is a trace of thermodynamic dimensional reduction from $D=3$ to $D=2$.
\end{abstract}
\date{\today }
\maketitle

\section{Introduction}

Heisenberg uncertainty principle is considered as a cornerstone of the quantum theory from which a new perspective of the physical world is realized. It addresses the fact that the position and momentum of a test particle cannot be measured with the same precision simultaneously. This feature results in a relevant algebra in which position and momentum are not commuting operators. In consequence, the phase space does not comprise the points with complete resolution. However, this revolutionary principle proposes that position and momentum can be measured with zero uncertainty separately, leading to infinite imprecision for conjugate quantity (operator). Quantum gravity theories have shed light on this issue by considering the impact of gravity on quantum theory~\cite{Hossenfelder2013}. Theories such as string theory, noncommutative geometry, doubly special relativity, and loop quantum gravity suggest a modified version of the Heisenberg uncertainty principle (HUP) to overcome this problem. The new modified HUP features a minimal measurable length at the scale of the Planck length, which results in the discontinuity of the spacetime manifold~\cite{Aguilar2013}. Also, a simple thought experiment of a micro black hole verifies this claim~\cite{Scardigli1999}. The impact of a minimal measurable length has been widely inspected in different phenomenological problems of physics (see for instance Ref.~\cite{Nozari2017,Khodadi2018} and references therein), and its validity is almost commonly accepted as a genuine feature of the very nature of the background spacetime at the Planck length scale. In advance, the symmetry of the phase space symplectic manifold makes it reasonable to have also a minimal measurable momentum which can be obtained from the curvature of the background spacetime~\cite{Hinrichsen1996,Mignemi2010}. If AdS/dS spacetime is considered as the background topology of the spacetime, which features curvature as proportional to the inverse squared AdS/dS radius, noncommutativity of the momentum space must also be regarded. As a result, HUP would be modified by a factor proportional to the cosmological constant, which impacts the physical phenomena at cosmological distances. Excitingly, this modified version of HUP affects the quantum world at the scale of energy far from which the quantum theory was derived. This modified version of HUP (known as Extended Uncertainty Principle (EUP)) reads as follows
$$\delta x \delta p \geq \frac{\hbar}{2}\big[1+\alpha (\delta x)^2\big]$$
where $\alpha$ denotes a factor that meaningfully influences the physical phenomena at very large distances. This relation leads to the noncommutativity of momentum space, $[p_{i},p_{j}]\neq 0$, and is the basis of the triply special relativity~\cite{Smolin2004}. See some efforts devoted to probing the effects of EUP on physical problems in Refs.~\cite{Mirza2009,Zho2009,Filho2016,Mureika2019,Nozari2019}.

One of the potential problems worth noting, which is regarded in the semi-classical manner, is the dynamics of a quantum bouncer that experiences the Newtonian gravity. The effects of a minimal measurable length on the spectrum of a quantum bouncer were firstly studied in Ref.~\cite{Nozari2011}. Then some important constraints on the quantum gravity parameter of a minimal length uncertainty relation were obtained in Ref.~\cite{Pedram2011}, where the spectrum of a system of ultracold neutrons was treated in a high energy regime. Here our goal is to study a system of quantum bouncers in a version of quantum mechanics that is viable in the very low energy regime, close to the zero temperature. This low energy modified quantum mechanics is equipped with a minimal measurable momentum as an infrared cutoff. This problem is important to be reconsidered because of its close relation to the famous Nesvizhevsky \emph{et al.} experiment~\cite{Nevi2002}. In fact, the problem of the semi-classical impact of gravity on the spectrum of a quantum particle has become noticeable after the Nesvizhevsky \emph{et al.} experiment, which regarded the real fingerprint of the Earth's gravity on the quantum states of ultracold neutrons in an experiment. They employed the very low energy neutrons bouncing on a soft floor, which are trapped by the gravity of the Earth. The important fact is that extremely weak gravity, in non-relativistic manner, compared with electromagnetic and nuclear forces, is here responsible for the creation of the quantum states. In this regard, the observation of the quantum states for the bouncing ultracold neutrons hit the scientific community and drew much attention, in advance. Now it seems to be very interesting to reconsider the problem with a low energy modified version of standard quantum mechanics, the issue that is the subject of the present study.

In this work, we are going to theoretically consider a low energy modified version of quantum mechanics to probe the behavior of quantum bouncers near-zero temperature. While it is usually believed that the issue of dimensional reduction is a high energy effect, here we seek for possibility of having a trace of thermodynamic dimensional reduction in this low energy regime. Some features of an IR-deformed quantum mechanics have been studied in recent years (see, for instance, ~\cite{Mirza2009}), and the subject is worth attention for a quantum bouncer problem due to its close relation with the spectrum of ultracold neutrons in the Earth's gravitational field. We probe the impression of the quantum gravity (via an extended uncertainty relation admitting a minimal measurable momentum) on the problem of a system of very low energy quantum bouncers. In other words, IR deformation in the spectrum of quantum bouncers, due to the low energy version of quantum gravity, is inspected in detail. If being successful, it would be exciting to realize the quantum gravity impacts on the low energy scale, which is normally considered irrelevant to any quantum behavior. This essentially provides a low energy scheme to test quantum gravitational effects in a table-top experiment.

We proceed with the problem in a one-dimensional case using the first-order WKB approximation, considering very low energy of the quantum particles, which is valid within the IR regime. Then we find the energy spectrum and the bound states by the use of modified Hamiltonian due to the IR-deformed algebra. Putting in the mass of neutrons, the energy levels of ultracold neutrons are calculated with the consideration of the IR-deformed HUP impression. Regarding the calculated energy levels, it is seen that observed energy levels claimed in Nesvizhevsky \emph{et al}. paper agree with the calculated results, with a negative constant deformation. Moreover, probability density functions of the bound states suggest a slight reduction of the spatial frequency, in comparison with undeformed case. Afterward, the thermodynamics of quantum bouncers is tackled by the use of ensemble theory. We calculate the modified entropy and internal energy for a one-dimensional IR-deformed quantum bouncer, that become stationary within the temperature under $1$ K. In advance, the three-dimensional case is regarded, and the same procedure is performed to reach the IR-deformed bound states. By the knowledge of the three-dimensional IR-deformed energy levels for a quantum bouncer, the corresponding thermodynamics problem is tackled. Entropy, internal energy, and heat capacity of a system of three-dimensional IR-deformed quantum bouncers are calculated and analyzed. The results excitingly imply the elimination of the stationary behavior under $1$ K, and the asymptotic trend suggests a thermodynamic dimensional reduction from $D=3$ to $D=2$ dimensions, in contrast to dimensional reduction from $D=3$ to $D=1$ in the high energy regime of quantum gravity modifications~\cite{Nozari2015,Carlip2017}. After all, asymptotic thermodynamic behavior of an IR-deformed quantum bouncer imitates the classical behavior of equipartition energy, while leaving the IR regime of energy. This work illuminates the approximate behavior of the quantum bouncers in very low energy regime with consideration of the infrared quantum gravity trace.

\section{IR deformation of one-dimensional quantum bouncers}

Dynamics of a quantum particle that is confined to the Newtonian gravity provides an amazing route towards identification of the impact of non-relativistic gravity on quantum states, i.e. the concept of a quantum bouncer. The resulting quantum bouncer was somehow a fabricated example of a linear one-dimensional potential, but a recent experiment of ultracold neutrons, with low energy neutrons, implied the possibility of quantum bouncers in the real world~\cite{Nevi2002}. The crucial point is that the creation of quantum states is handled by the gravitational field of the Earth. Then this problem can be considered as one features a fingerprint of semi-classical gravitation in the quantum world. Here, we attempt to modify the well-known Airy equation of a quantum bouncer within an IR-deformed Heisenberg algebra. A quantum bouncer model can be established in the lab by providing a very low energy regime of a quantum particle. Reflection from a barricade can happen for a quantum particle if its energy is very low enough that it can not transmit the barrier, then is reflected. This problem is presented so far within ordinary quantum mechanics by well-known Airy equation~\cite{valee2004}. However, as mentioned above, the authentic version of quantum mechanics for a quantum bouncer seems to be the one corrected in the realm of IR energy. To solve this problem, we incorporate modification of the standard Heisenberg algebra due to the existence of a minimal measurable momentum. Therefore, the canonical commutation relation is altered due to the extended uncertainty principle (EUP) as follows
$$X=x \,,\,\quad\quad P=(1+\alpha x^{2})p\,,$$
resulting in
\begin{equation}
[X,P]=i\hbar(1+\alpha x^{2})\,\,\, \Rightarrow \,\,\, \delta X \delta P \geq \frac{\hbar}{2}(1+ \alpha X^{2})\,.
\end{equation}
Then the one-dimensional Hamiltonian operator reads
\begin{equation}
  \hat{H} = \dfrac{\hat{P}^{2}}{2m}+V(\hat{X})=\frac{-\hbar^{2}}{2m}\Big(1+\alpha X^{2}\Big)^{2}\,\frac{d^{2}}{dX^{2}}+mgX\,.
\end{equation}
In result, IR-corrected Schr\"{o}dinger equation for stationary states comes out ($X=x$)
\begin{equation}\label{1}
  -\frac{\hbar^{2}}{2m}(1+\alpha x^{2})^{2}\,\frac{d^{2}\Psi(x)}{dx^{2}}+mgx\,\Psi(x)=E\Psi(x) .
\end{equation}
We then employ a point canonical transform $\Psi(x)=G(x)\,\Phi(u)$, where $u$ is a function of $x$, to convert Eq. (\ref{1}) to a standard Schr\"{o}dinger equation in advance with an effective potential as $\mathcal{V}(u)$ instead of Newtonian potential. Calculations come up with the result as follows
\begin{align}
& u(x)=\frac{1}{\sqrt{\alpha}}\,\arctan{\sqrt{\alpha}\,x}  \\
& G(x)=\sqrt{1+\alpha\,x^2}\,,
\end{align}
with $\mathcal{V}(u)$ and $\mathcal{E}$ as below
\begin{align}
& \mathcal{V}(u)=\frac{mg}{\sqrt{\alpha}}\,\tan{\sqrt{\alpha}\,u}\quad 0<u<\frac{\pi}{2} \\
& \mathcal{E}=E+\frac{\alpha\,\hbar^2}{2m}\,.\\
\end{align}
It finally leads to the following standard Schr\"{o}dinger equation with new canonical variables $u$ and its momentum $P_u$
\begin{equation}\label{2}
-\frac{\hbar^{2}}{2m}\frac{d^{2}\Phi(u)}{du^{2}}+\mathcal{V}(u)\,\Phi(u)=\mathcal{E}\Phi(u)\,.
\end{equation}
It is obvious that $\mathcal{V}(u)$ tends to infinity for both $u=0\,,\,\frac{\pi}{2\,\sqrt{\alpha}}$ as an impenetrable barrier is set at $x=0$\,. Thus, $\mathcal{V}(u)$ only admits bound states, and there would not be any possibility to have scattering solution to Eq. (\ref{2}), with the same analysis for $\Psi(x)$ as the solution to Eq. (\ref{1}). Going forward, we proceed with solving Eq. (\ref{2}) using first-order WKB approximation in the classical region, as implied by the resulting bound states, where $\mathcal{E}$, say energy of the particle subject to the effective potential, is greater than the imposed potential $\mathcal{V}(u)$. In this way, turning point can be easily calculated as $u_t=\frac{1}{\sqrt{\alpha}}\,\arctan{\frac{\mathcal{E}\,\sqrt{\alpha}}{mg}}$\,. In advance, we note that $\mathcal{E}$ is near zero in the IR realm of the ultracold quantum bouncers. Thus turning point becomes $\frac{\mathcal{E}}{mg}$, still dependent on deforming parameter $\alpha$, and effective potential $\mathcal{V}(u)$ reads
\begin{align}
& \mathcal{V}(u)=mgu \\
& u_t=\frac{\mathcal{E}}{mg}\quad \text{where}\quad \mathcal{E}=E+\frac{\alpha\,\hbar^2}{2m}\,.
\end{align}
Using this asymptotic behaviour, the semi-classical WKB solution in the IR realm for the bound states $\Phi(u)$ is acquired as follows
\begin{equation}\label{3}
\Phi(u)\approx\frac{2\,A}{\sqrt{2m^2g\,(u_t-u)}}\,\sin\left({\frac{\sqrt{2m^2g}}{\hbar}}\,\int_{u}^{u_t=\frac{\mathcal{E}}{mg}}\sqrt{2m^2g\,\left(u^{\prime}_t-u^{\prime}\right)}\,\,du^{\prime}+\frac{\pi}{4}\right)\,,
\end{equation}
with $A$ as a normalization constant. Exploiting boundary condition at $u=0$, Eq. (\ref{3}) amounts to the quantization rule as the following
\begin{equation}\label{4}
\frac{\sqrt{8m^2g}}{3\hbar}\,u_t^{\,\frac{3}{2}}\,+\frac{\pi}{4}=n\pi\,.
\end{equation}
Setting $u_t=\frac{\mathcal{E}}{mg}$, bound states energies as dependent on the deforming parameter $\alpha$ are obtained as below
\begin{equation}\label{5}
E_n=\frac{m^{\frac{1}{3}}\,g^{\frac{2}{3}}}{2}\,\left[3\pi\hbar\left(n-\frac{1}{4}\right)\right]^{\frac{2}{3}}-\frac{\alpha\,\hbar^2}{2m}\quad n=1,2,3,...\,\,.
\end{equation}
It is interesting to note that Eq. (\ref{5}) verifies the undeformed spectrum of quantum bouncers, say $\alpha=0$, as eigenvalues for the Airy equation. Setting $\alpha=10^{10}$, we compare the resulting deformed spectrum of ultracold neutrons with the undeformed case for the first ten bound states as follows in Table. I. We also note that the deformation term in Eq. (\ref{5}) has a constant negative value for all bound states as acquired with our approach. Noteworthy to be added, improving Nesvizhevsky \emph{et al}. experiment with power to measure the excited states of ultracold neutrons can certainly lead to approaximating the deformation parameter $\alpha$ in a phenomenological manner, using our result. But in this work, we are more involved with the impact of IR modification on the thermodynamic behavior of quantum bouncers though.

\begin{table}[ht!]
  \centering
\begin{tabular}{cccc}
\noalign{\smallskip}\hline \hline \noalign{\smallskip}
Energy Level    &    Deformed UCN   &    Undeformed UCN \\
\noalign{\smallskip} \hline \noalign{\smallskip}
1   &   1.18 peV  &   1.39 peV \\

2   &    2.24 peV   &    2.45 peV \\

3   &   3.11 peV    &   3.31 peV \\

4   &   3.87 peV    &   4.08 peV \\

5   &   4.57 peV    &   4.77 peV \\

6   &   5.21 peV    &   5.42 peV \\

7   &   5.83 peV    &   6.03 peV \\

8   &   6.41 peV    &   6.62 peV \\

9   &   6.97 peV    &   7.17 peV \\

10   &   7.50 peV    &   7.71 peV \\

\noalign{\smallskip} \hline \hline \noalign{\smallskip}
\end{tabular}
  \caption{ \small{Calculated deformed spectrum of ultracold neutrons for the first 10 bound states in comparison with undeormed case for $\alpha=10^{10}$.}}
\end{table}

In the next step, we investigate the IR-deformed WKB wave functions to find any possible contrasting behavior in comparison with the undeformed case. In this way, $u\approx x$ in the IR realm and Eq. (\ref{3}) reads
\begin{equation}\label{6}
\Phi(x)\approx\frac{2\,A}{\sqrt{2m^2g\,(\frac{\mathcal{E}}{mg}-x)}}\,\sin\left({\frac{\sqrt{2m^2g}}{\hbar}}\,\int_{x}^{\frac{\mathcal{E}}{mg}}\sqrt{2m^2g\,\left(\frac{\mathcal{E}}{mg}-x^{\prime}\right)}\,\,dx^{\prime}+\frac{\pi}{4}\right)\,.
\end{equation}
Rescaling $m=2\,,\,g=\frac{1}{2}\,,\,\text{and}\,\hbar=1$, IR-deformed WKB solution to Eq. (\ref{1}) becomes
\begin{equation}\label{7}
\Psi(x)\approx\frac{2\,A\,\sqrt{1+\alpha\,x^2}}{\left(\mathcal{E}-x\right)^{\frac{1}{4}}}\,\sin\left[\frac{2}{3}\left(\mathcal{E}-x\right)^{\frac{3}{2}}+\frac{\pi}{4}\right]\,,
\end{equation}
where $\mathcal{E}=\left[\frac{3\,\pi}{2}\,\left(n-\frac{1}{4}\right)\right]^{\frac{2}{3}}$ for  $n=1,2,3,...\,\,$. The deformed probability density functions based on Eq. (\ref{7}) are depicted in comparison with undeformed case for the second, fourth, sixth, and eighth bound states in FIG. 1. Looking at the FIG. 1, two points can be identified. First, a slight movement of the deformed solution towards right side can be recognized, e.g. the first root of the deformed case is $1.76$ versus $1.73$ for that of the undeformed case. This fact accordingly insinuates the less spatial frequency for the wave functions of IR-deformed quantum bouncers, which is acceptable, and expectable, because of the very low energy realm of quantum gravity modifications due to EUP algebra. Second, we explicitly see the more localization of the deformed particle by increasing the deformation parameter, thus more probability for the same spatial interval in the peak region of any deformed bound state, as is seen in FIG. 2. This is also expectable in the IR regime of energy in advance since the reduced momentum of a deformed quantum bouncer in the IR energy regime would increase the probability of finding in the peak spatial region, i.e. moving towards the semi-classical behavior of a quantum particle.
\begin{figure}[ht!]
\centering
\includegraphics[width=7.8cm]{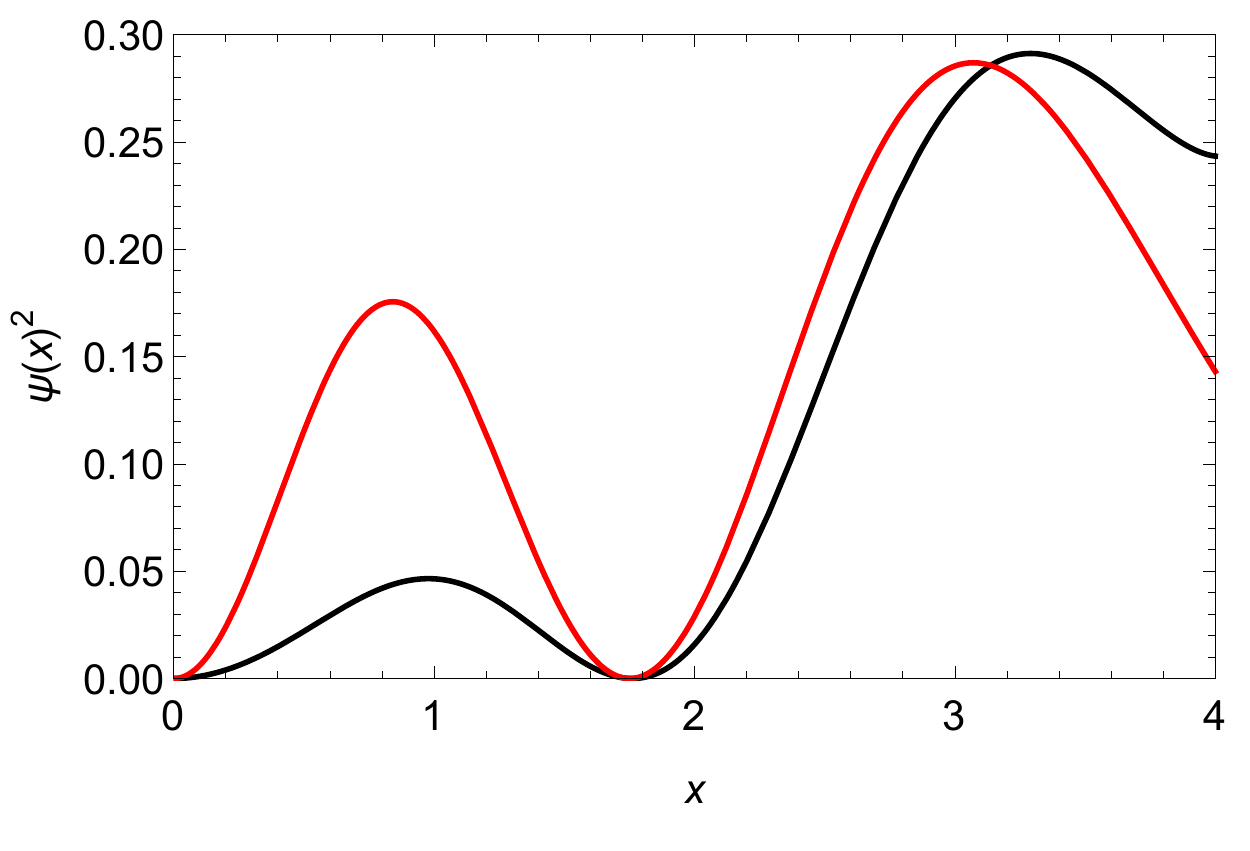}
\includegraphics[width=7.8cm]{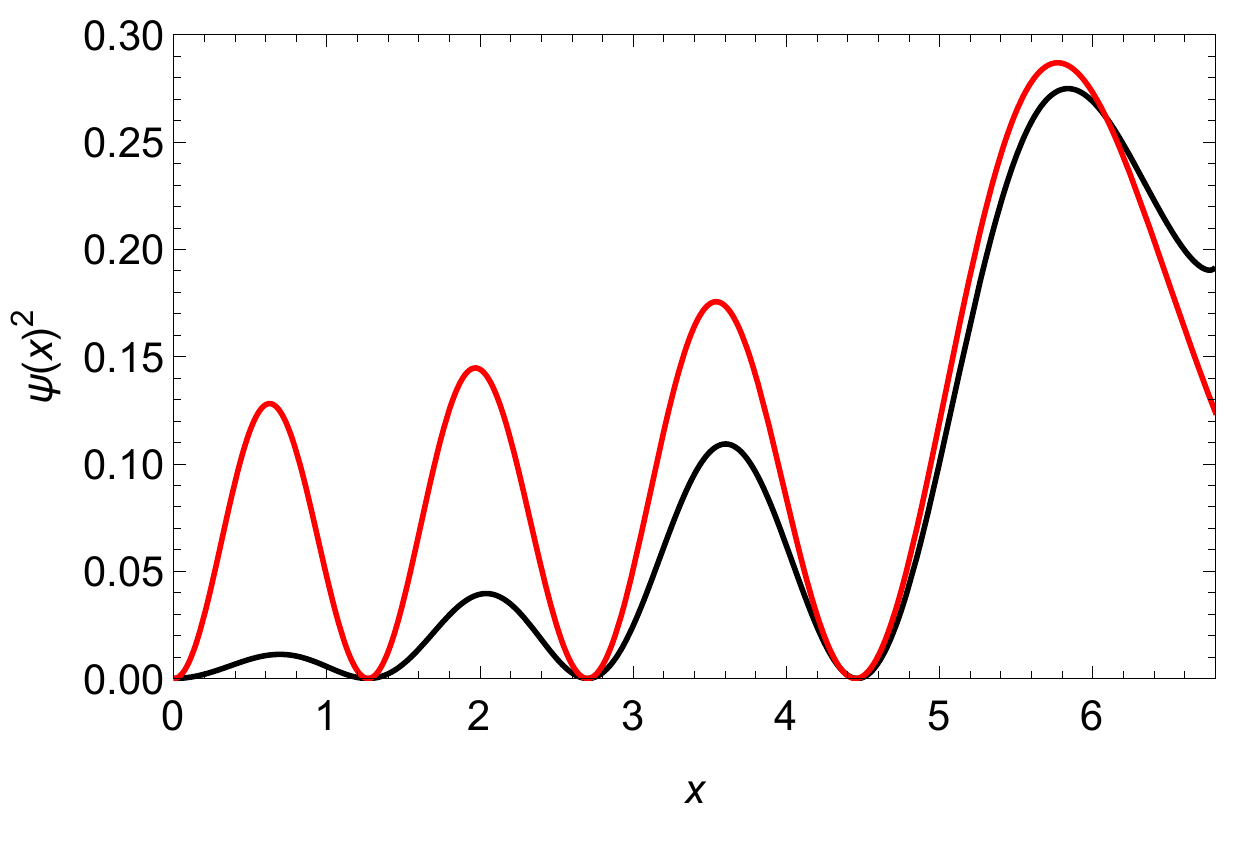}
\includegraphics[width=7.8cm]{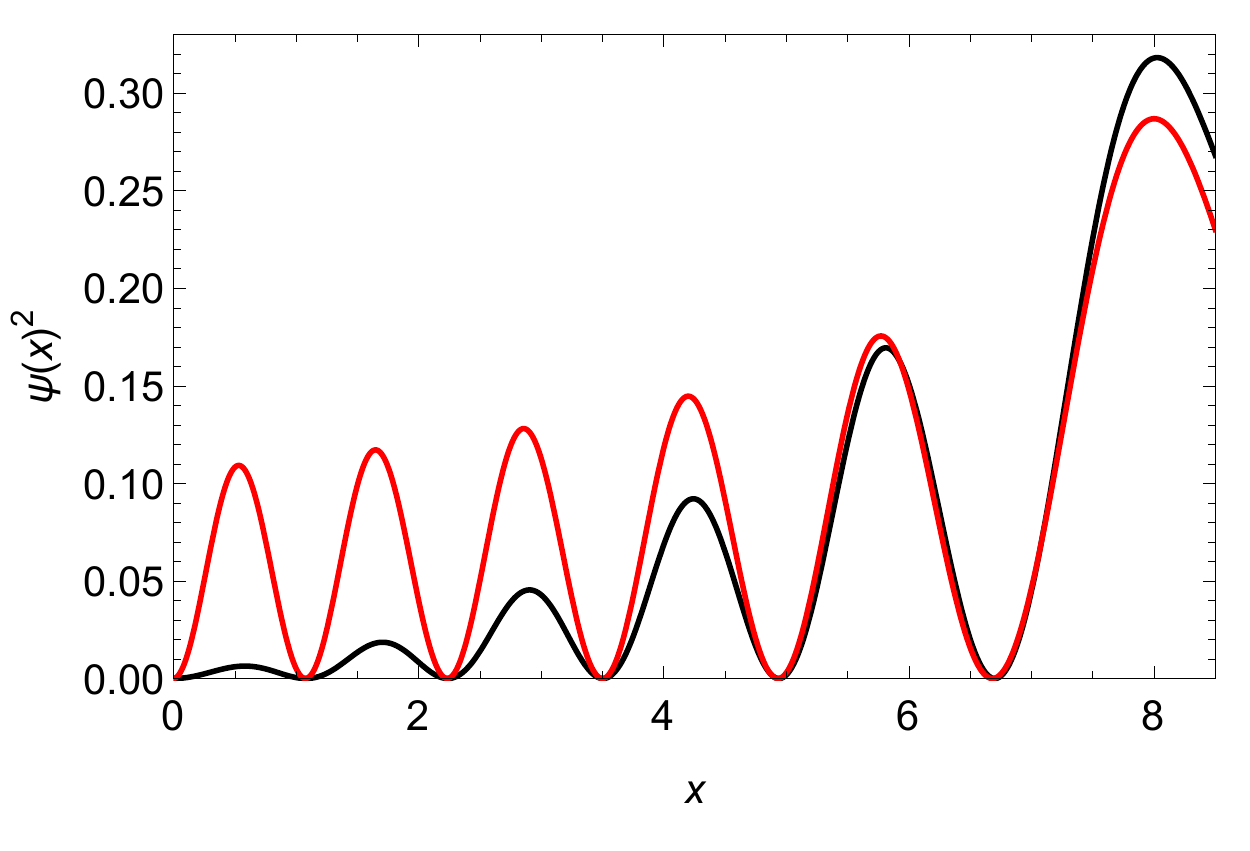}
\includegraphics[width=7.8cm]{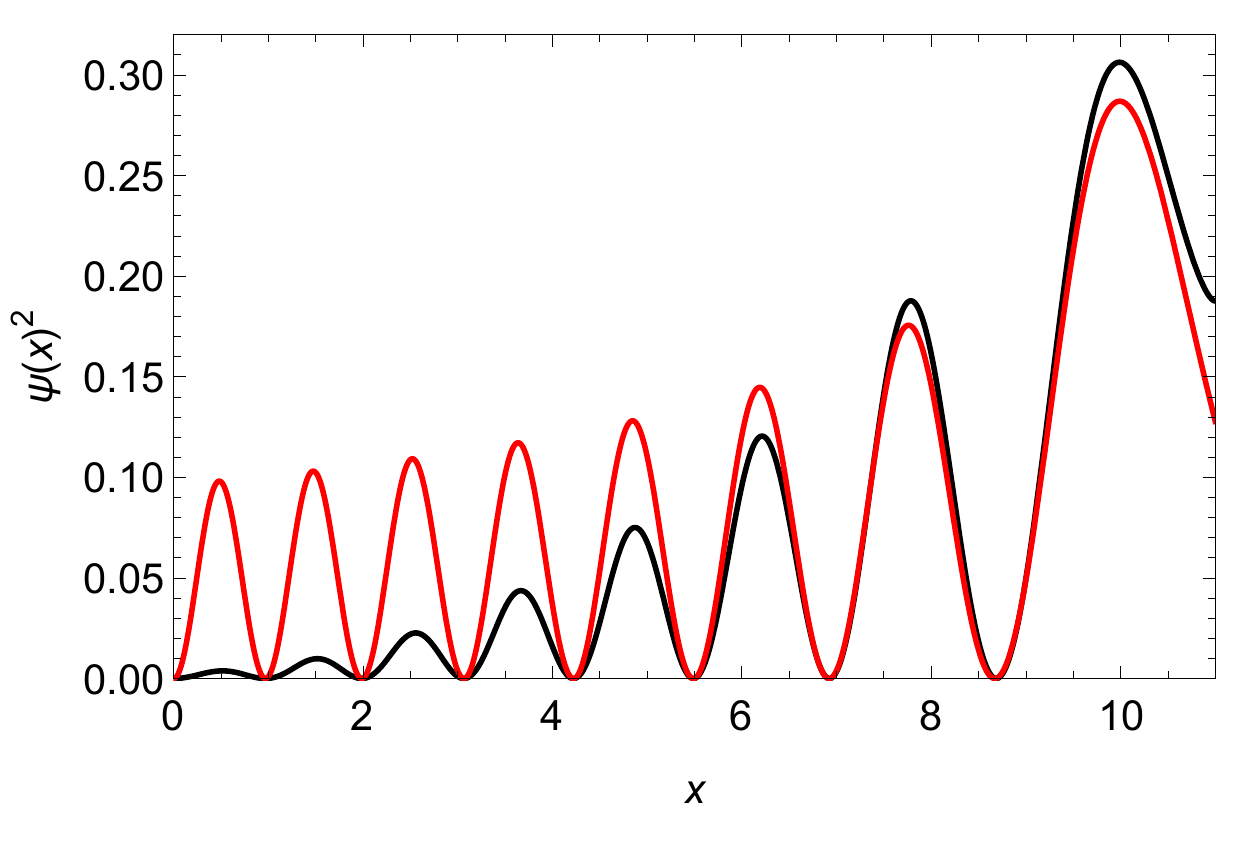}
\caption{\small {The unnormalized  probability density functions of both IR-deformed (black) and ordinary (red) quantum bouncers for second state (top left), fourth state (top right), sixth state (bottom left), and eigth state (bottom right) setting $\alpha = 1$.}}
\end{figure}

\begin{figure}[ht!]
\centering
\includegraphics[width=14cm]{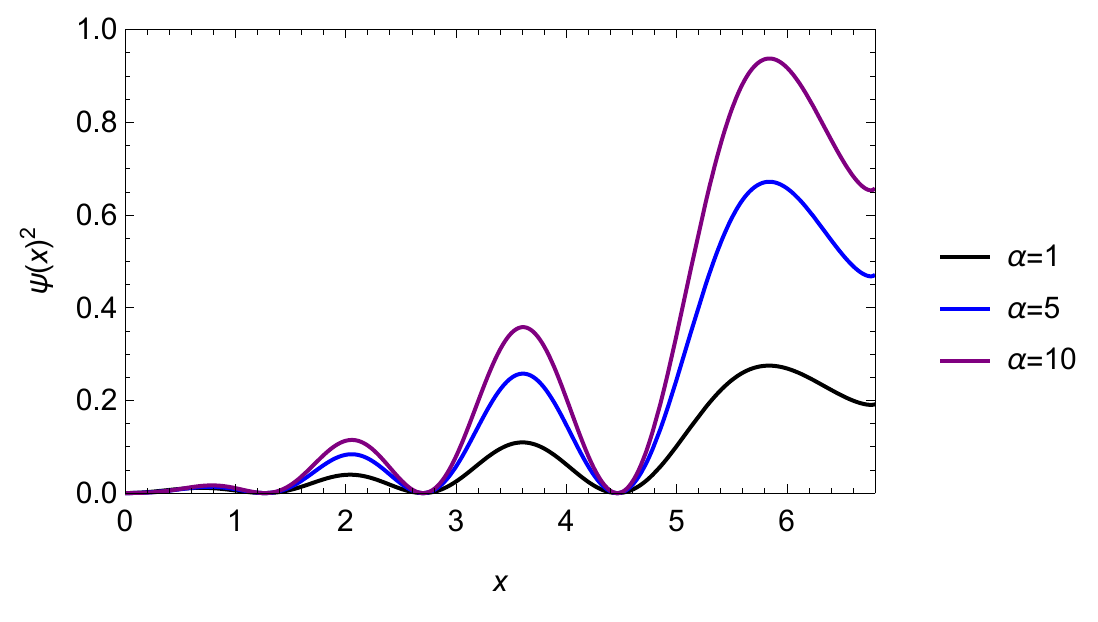}
\caption{\small {Comparison of IR-deformed probability density function for the fourth state as impacted by three different values of the deforming parameter, $\alpha = 1,5,10$.}}
\end{figure}

\section{One-dimensional IR-deformed thermodynamics}

With the deformed energy levels as in Eq. (\ref{5}), we can proceed with the problem of IR correction to treat the thermodynamics of a quantum bouncer. All the thermodynamic properties can be derived using the partition function, which can be acquired with IR-deformed energy levels as follows
\begin{equation}
  Q(\beta)=\sum_{n=1}^{\infty} e^{- \beta (E_{_{n}})_{_{IR}}}\,, \quad\quad\quad \beta = \frac{1}{k_{B}T}\,,
\end{equation}
where $k_{B}$ is the Boltzmann constant and $T$ is temperature. As usual, our main focus is on IR regime where the temperature is low. In this low temperature regime, $\beta$ is larger than the high temperature case. Therefore, in order to have considerable contributions in $e^{- \beta (E_{_{n}})_{_{IR}}}$, one has to consider low energy cases of the problem, which are the cases with small quantum number, say $n$. Moreover, a quantum bouncer behaves semiclassically in IR regime, then low energy terms are more closely packed together. Considering these facts, the above-mentioned sum can be approximated with an integral over the IR-deformed energy levels as
$$Q(\beta)\approx \int_{1}^{\infty} e^{- \beta (E_{_{n}})_{_{IR}}} \ dn\,,$$
that is
\begin{equation}
Q(\beta)\approx \int_{1}^{\infty} e^{- \beta\Big[\Big(\frac{3 \pi}{2}(n-\frac{1}{4})\Big)^\frac{2}{3}-\alpha\Big]} \ dn \ .
\end{equation}
But, we decide to separate the ground state as it results in the dominant term in the case of high $\beta$, therefore leaving it as a single term would increase the precission of converting the sum to the integral. So, we rewrite the partition function as follows
\begin{equation}
  Q(\beta)\approx e^{\alpha\beta} \left(e^{-2.34\,\beta}+\int_{2}^{\infty} e^{- \beta\left[\frac{3\,\pi}{2}\,\left(n-\frac{1}{4}\right)\right]^\frac{2}{3}} \ dn\right)\,.
\end{equation}

At this point, we have to evaluate the integral to reach a mathematical relation for the partition function. This integral can be approximated numerically by nonlinear modeling with a chosen function of desire, using the \emph{least square approach}. We choose this function as $a \beta ^{-\frac{1}{2}} e^{-b \beta}$ \ with $a \ \text{and} \ b$ as the constants to be computed. The result is $a=5.30 \ \text{and} \ b=8.52$ \ with $0.0003$ as the standard deviation of the difference between true and approximated data in the range of very low energy regime, say $ T \ \text{within} \ 0.01 \ \text{to} \ 1$. Finally, the IR-deformed partition function of a quantum bouncer reads
\begin{equation}\label{7.5}
  Q(\beta)_{_{IR}}\approx e^{\alpha\beta}\left(e^{-2.34\,\beta}+5.30 \, \beta ^{-\frac{1}{2}}e^{-8.52 \, \beta}\right).
\end{equation}
To explore the thermodynamic behavior of a physical system of quantum bouncers, we employ canonical ensemble theory and gather up $N$ indistinguishable quantum bouncers with the partition function $Q_{N}(\beta)=\frac {[Q_{1}(\beta)]^{N}}{N!}$\,. Setting $k_{B}=1$, Helmholtz free energy defined as $ A(N,T)\equiv-T \ln Q_{N}(T)$ can be calculated as
\begin{equation}
A(N,T)\approx T\big(N \ln N -N \big)-N\alpha-NT \ln \left( e^{-\frac{2.34}{T}}+5.30\,T ^{\frac{1}{2}}e^{-\frac{8.52}{T}}\right)\,,
\end{equation}
where we have used Stirling's approximation $\ln N!\approx N\ln N-N$. Then entropy of the ensemble, defined as $S(N,T)\equiv-\big(\frac{\partial A}{\partial T}\big)_{N}$, can be obtained as follows
\begin{align}
S(N,T)= N \Bigg(1&+ \ln \left[e^{-\frac{2.34}{T}}+5.30\,T^{\frac{1}{2}}e^{-\frac{8.52}{T}}\right] \\ \notag &+\frac{\frac{2.34}{T}\,e^{-\frac{2.34}{T}}+e^{-\frac{8.52}{T}}(2.65\,T^{\frac{1}{2}}+45.15\,T^{-\frac{1}{2}})}{ e^{-\frac{2.34}{T}}+5.30\,T ^{\frac{1}{2}}e^{-\frac{8.52}{T}}}-\ln N \Bigg) \ .
\end{align}
In result, the entropy of a quantum bouncer acquired from the IR-deformed partition function reads
\begin{align}
  S(N=1,T)=1+ \ln \left[e^{-\frac{2.34}{T}}+5.30\,T^{\frac{1}{2}}e^{-\frac{8.52}{T}}\right]+\frac{\frac{2.34}{T}\,e^{-\frac{2.34}{T}}+e^{-\frac{8.52}{T}}(2.65\,T^{\frac{1}{2}}+45.15\,T^{-\frac{1}{2}})}{ e^{-\frac{2.34}{T}}+5.30\,T ^{\frac{1}{2}}e^{-\frac{8.52}{T}}}
  \end{align}
which is depicted in FIG .3.
\begin{figure}[ht!]
\centering
\includegraphics[width=12cm]{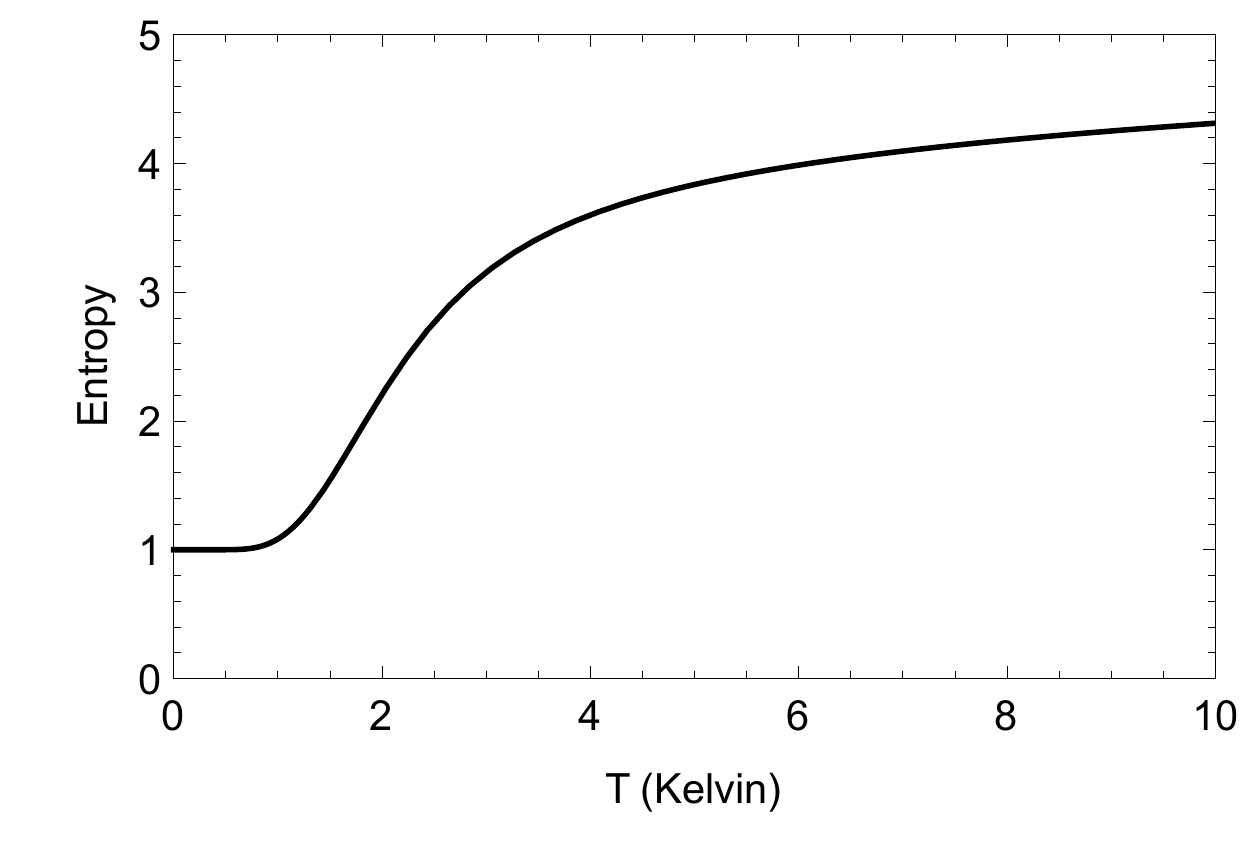}
\caption{\small {Entropy of an IR-deformed quantum bouncer in the low temperature regime. Apparently the particle reaches a minimum non-zero entropy for $T<1$\,K.  }}
\end{figure}
Asymptotic behavior of the entropy of an IR-deformed quantum bouncer is as\, $$\lim_{_{\,T \to 0}} \,S(T)=1\,,$$ $$\lim_{_{\,T \to \infty}} \,S(T)= 1+2.65\,\sqrt{T}+\ln(1+5.30\,\sqrt{T})\approx\,2.65\,\sqrt{T}\,.$$
Next step is to calculate the internal energy of an ensemble of IR-deformed quantum bouncers using the relation
$U(N,T)\approx T^{2}\,\frac{\partial \ln Q_{N}(T)}{\partial T}$
\begin{equation}
\frac{U(N,T)}{N}=-\alpha+\frac{2.34\,e^{-\frac{2.34}{T}}+e^{-\frac{8.52}{T}}(2.65\,T^{\frac{3}{2}}+45.15\,T^{\frac{1}{2}})}{e^{-\frac{2.34}{T}}+5.30\, T^{\frac{1}{2}}e^{-\frac{8.52}{T}}}\,.
\end{equation}
The behavior of mean internal energy of an IR-deformed quantum bouncer is shown in FIG. 4, implying some interesting features.
\begin{figure}[ht!]
\centering
\includegraphics[width=16cm]{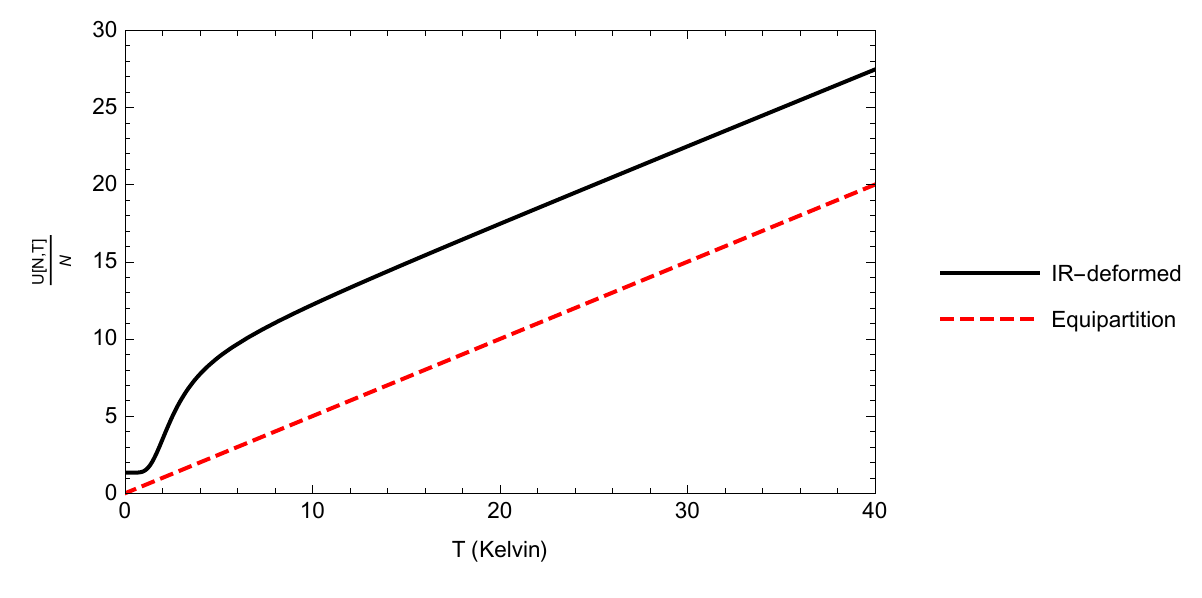}
\caption{\small {The mean internal energy of an IR-deformed quantum bouncer as compared with classical equipartition energy of $\frac{1}{2}\,T$ (with $k_{B}=1$) for the case $\alpha=1$. It is seen that minimal mean internal energy of an IR-deformed quantum bouncer is not zero and is equal to the IR-deformed ground state energy of a quantum bouncer. }}
\end{figure}
Considering asymptotic behavior of $\frac{U(N,T)}{N}$, we find that it tends to the classical equipartition energy of $\frac{1}{2}\,T$ as temperature rises to high energy regime, that is,  $\lim_{_{\,T \to \infty}}\,\frac{U(N,T)}{N}\approx\frac{1}{2}\,T$\,. This feature, taking into account the correspondence principle, shows the credibility of the acquired thermodynamic results. Moreover, as temperature decreases to IR realm, the mean internal energy goes towards the IR-deformed ground state of quantum bouncers as expected, $\lim_{_{\,T \to 0}} \,\frac{U(N,T)}{N}=2.34-\alpha$\,. It can be acceptable to contemplate quantum bouncers residing in their ground state while temperature goes to zero, which shows the non-zero minimal energy of IR-deformed quantum bouncers in contrast to the classical bouncers.

\section{Generalization to three-dimensions}
To tackle the problem of three-dimensional quantum bouncer with IR-deformed Hamiltonian, we assume the overall wave function of the quantum particle to be separable, i.e. $\Psi(x,y,z)=\psi(x)\,\phi(y)\,\theta(z)$\,. Moreover, we contemplate a particle inside a three-dimensional infinite square well of the size $L$ in both $x$ and $y$ directions while experiencing gravity of the Earth in the $z$ direction. Then the relevant potential takes the following form (note that we have rescaled $m=\frac{1}{2}\,,\,g=2\,,\,\hbar=1$ for simplicity)
\begin{equation}
V(x,y,z)=
\begin{cases}
  z & 0<x<L ,\,\, 0<y<L \,,\,\, z>0 \\
  \infty &  \text{elsewhere}\,.
\end{cases}
\end{equation}
To proceed, we consider the IR-deformed Heisenberg algebra as $X_{i}=x_{i}$ and $P_{i}=(1+\alpha x_{i}^2)p_{i}$ where $i=1,2,3$. Then the modified three-dimensional Hamiltonian of an IR-deformed quantum bouncer reads
$$H_{_{IR}}=P^{2}+V=-\Big[(1+\alpha x^{2})^{2}\,\frac{\partial^{2}}{\partial x^{2}}+(1+\alpha y^{2})^{2}\frac{\partial^{2}}{\partial y^{2}}+(1+\alpha z^{2})^{2}\frac{\partial^{2}}{\partial z^{2}}\Big]+V(x,y,z)\,.$$

The next step is to write the modified time-independent Schr\"{o}dinger equation for probing the IR-deformed three-dimensional bound states. The equation is simplified after some calculations as
\begin{multline}\label{8}
  (P^{2}+V)\Psi=E\Psi \\
  \Rightarrow  -\bigg[\frac{(1+\alpha x^{2})^{2}\,\frac{d^{2}\psi(x)}{d x^{2}}}{\psi(x)}+\frac{(1+\alpha y^{2})^{2}\,\frac{d^{2}\phi(y)}{d y^{2}}}{\phi(y)}+\frac{(1+\alpha z^{2})^{2}\,\frac{d^{2}\theta(z)}{d z^{2}}}{\theta(z)}\bigg]+z=E\,.
\end{multline}
 Since commutativity of space is assumed to be valid in the case of IR-deformed Heisenberg algebra, it is acceptable to claim that $x$, $y$, and $z$ can independently vary without impact on each other to finally result in a constant $E$. In consequence, each independent part of the above equation must be set equal to a constant. Therefore, Eq. (\ref{8}) can be cast into three independent equations as follows
 \begin{align}
   &-\frac{(1+\alpha x^{2})^{2}}{\psi(x)}\,\frac{d^{2}\psi(x)}{d x^{2}}=E_{x}\label{9}\\
   &-\frac{(1+\alpha y^{2})^{2}}{\phi(y)}\,\frac{d^{2}\phi(y)}{d y^{2}}=E_{y}\label{10} \\
   &-\frac{(1+\alpha z^{2})^{2}}{\theta(z)}\,\frac{d^{2}\theta(z)}{d z^{2}}+z=E_{z} \label{11}\\
   &E=E_{x}+E_{y}+E_{z} \, .
 \end{align}
Eq. (\ref{11}) was solved previously for the case of one-dimensional quantum bouncer, and $E_{z}$ was obtained as discussed before. Then, by solving two Eqs. (\ref{9}) and (\ref{10}), the IR-deformed energy levels will be established, resulting in the three-dimensional IR-deformed partition function. To proceed, the present task is to calculate $E_{x}$ and $E_{y}$, which feature the same mathematical relation. We employ the point canonical transform similar to the one-dimensional case, i.e. $\psi(x)=\mathcal{G}(x)\mathcal{F}(v(x))$, for solving Eq. (\ref{9}). Accordingly, $\psi(x)$ can then be acquired easily without any approximation as follows
\begin{align}
  \psi(x)=C_{1}\sqrt{1+\alpha x^{2}}\,&\exp\bigg\{i\Big(\sqrt{1+\dfrac{E_{x}}{\alpha}}\Big)\,\arctan(\sqrt{\alpha}\,x)\bigg\}\notag \\
  &+iC_{2}\frac{\sqrt{1+\alpha x^{2}}\,\exp\bigg\{-i\Big(\sqrt{1+\dfrac{E_{x}}{\alpha}}\Big)\,\arctan(\sqrt{\alpha}\,x)\bigg\}}{2\sqrt{E_{x}+\alpha}}\,.
\end{align}
To move further, boundary conditions as $\lim_{\,x \to 0}\psi(x)=0$\, and $\lim_{\,x \to L}\psi(x)=0$\, must be imposed. First, considering $\lim_{\,x \to 0}\psi(x)=0$\, we reach
\begin{equation}\label{11.5}
\frac{C_{1}}{C_{2}}=\frac{-i}{2\sqrt{E_{x}+\alpha}}\,.
\end{equation}
Next, utilizing $\lim_{\,x \to L}\psi(x)=0$\, with the aid of Eq. (\ref{11.5}), we find
\begin{equation}\label{12}
C_{2}\bigg\{\Big(\sqrt{\frac{1+\alpha L^{2}}{E_{x}+\alpha}}\Big)\sin\Big[\Big(\sqrt{1+\frac{E_{x}}{\alpha}}\Big)\arctan(\sqrt{\alpha}\,L)\Big]\bigg\}=0
\end{equation}
in which $C_{2}$\, can not be zero as it leads to a nonsense solution. Therefore, Eq. (\ref{12}) necessitates the sine function to be zero. In result, the argument of the sine function must be equal to $n_{x}\,\pi$\, with $n_{x}=1,2,3,...$\,($n_{x}$ can not be zero as $L\neq 0$). Eventually it leads to
\begin{equation}\label{13}
E_{x}=\alpha\bigg[\bigg(\frac{n_{x}\,\pi}{\arctan(L\sqrt{\alpha})}\bigg)^{2}-1\bigg]\,,\,\,\,\,\,\, n_{x}=1,2,3,\ldots
\end{equation}
and in the same way, $E_{y}$\, is obtained as
\begin{equation}\label{14}
E_{y}=\alpha\bigg[\bigg(\frac{n_{y}\,\pi}{\arctan(L\sqrt{\alpha})}\bigg)^{2}-1\bigg]\,,\,\,\,\,\,\, n_{y}=1,2,3,\ldots\,.
\end{equation}
We need not worry about the denominator of Eqs. (\ref{13}) and (\ref{14}), since $L$\, and \,$\alpha$\, are not zero. To probe the rigor of these relations, asymptotic behavior must be inspected. If $L$ tends to be very large, as $L \to \infty$\,, we reach $E_{x}=\alpha\big(4n_{x}^{2}-1\big) \, (n_{x}=1,2,3,\ldots$)\,with the nonzero ground state of\, $(E_{x})_{\,min}=3\alpha$\,, which is in accordance with the impact of IR modification in the energy spectrum of a free particle. To be added, in the case of $\alpha \to 0$\,, it is expectedly reasonable to reach the undeformed spectrum of a particle within an infinite square well of size $L$\,. In this way, we acquire $\lim_{\,\alpha \to 0} E_{x}=\frac{n_{x}^{2}\,\pi^{2}}{L^{2}}$\, in consent with the claim above.

After all, the energy levels of a three-dimensional IR-deformed quantum bouncer are calculated using Eq. (\ref{5}) as
\begin{align}
E_{\,_{IR}}\approx\,\alpha&\bigg[\bigg(\frac{n_{x}\,\pi}{\arctan(L\sqrt{\alpha})}\bigg)^{2}+\bigg(\frac{n_{y}\,\pi}{\arctan(L\sqrt{\alpha})}\bigg)^{2}-2\bigg]
+\Big[\frac{3 \pi}{2}(n_{z}-\frac{1}{4})\Big]^\frac{2}{3}-\alpha\,\,.
\end{align}
Moving towards a more realistic case, we adopt very large $L$\, in order to have a quantum bouncer which would physically act like a free particle in $x$\, and $y$\, directions while being subject to the gravity of the Earth in $z$\, direction.
\begin{align}\label{15}
\lim_{\,L \to \infty} E_{\,_{IR}}\approx 4\alpha&\bigg[n_{x}^{2}+n_{y}^{2}-\frac{1}{2}\bigg]+\Big[\frac{3 \pi}{2}(n_{z}-\frac{1}{4})\Big]^\frac{2}{3}-\alpha
\end{align}

$$n_{x}\, , n_{y}\, ,n_{z}=1,2,3,\ldots\quad .$$
With the knowledge of IR-deformed energy levels for the three-dimensional case, thermodynamic behavior of an IR-deformed quantum bouncer can be tackled. Then, the partition function is written as follows
$$Q(\beta)=\sum_{n_{x},n_{y},n_{z}=1}^{\infty} e^{-\beta(E_{n_{x}}+E_{n_{y}}+E_{n_{z}})}\,,$$
and can be rewritten as
\begin{equation}
Q(\beta)\approx \int_{1}^{\infty} e^{- \beta E_{n_{x}}} \ dn_{x} \, \int_{1}^{\infty} e^{- \beta E_{n_{y}}} \ d n_{y} \, \int_{1}^{\infty} e^{- \beta E_{n_{z}}} \ d n_{z}\,.
\end{equation}
Putting in $E_{n_{x}}$\,, $E_{n_{y}}$\,, and $E_{n_{z}}$ calculated in Eq. (\ref{15}) and using Eq. (\ref{7.5}) as indicated in one-dimensional case,  it reads
$$
Q(\beta)_{_{IR}}\approx e^{\alpha\beta}\,\Big[\int_{1}^{\infty} e^{- 4\alpha \beta (n_{x}^{2}-\frac{1}{4})} \ dn_{x}\,\int_{1}^{\infty} e^{-4 \alpha \beta (n_{y}^{2}-\frac{1}{4})} \ dn_{y}\Big]\,\Big(e^{-2.34 \beta}+5.30\,\beta ^{-\frac{1}{2}}e^{-8.52 \beta}\Big)\,,
$$
which gives
\begin{equation}
Q(\beta)_{_{IR}}\approx e^{\alpha\beta}\,\Big[\int_{1}^{\infty} e^{- 4\alpha \beta (n_{x}^{2}-\frac{1}{4})} \ dn_{x}\Big]^{2}\Big(e^{-2.34 \beta}+5.30\,\beta ^{-\frac{1}{2}}e^{-8.52 \beta}\Big) \,.
\end{equation}
After evaluation of the integral within the bracket, $Q(\beta)$\, for a three-dimensional IR-deformed quantum bouncer takes the following form
\begin{equation}
Q(\beta)_{_{IR}} \approx \frac{\pi}{16\alpha}\,\beta^{-1}e^{3\alpha\beta}\Big[1-erf(2\sqrt{\alpha}\,\beta^{\frac{1}{2}})\Big]^{2}\Big(e^{-2.34 \beta}+5.30 \beta ^{-\frac{1}{2}}e^{-8.52 \beta}\Big) \,.
\end{equation}
It results in $Q_{N}(\beta)=\dfrac{[Q(\beta)]^{N}}{N!}$\, for a canonical ensemble of $N$ indistinguishable IR-deformed quantum bouncers for three-dimensional case. Afterwards, the thermodynamics can be acquired using standard relations. After calculation of the Helmholtz free energy, entropy of the ensemble is obtained as follows
\begin{align}
S(N,T)=&N\Big[2+\ln\big(\frac{\pi}{16\alpha}\big)+\ln\Big(\big[Te^{\frac{-2.34}{T}}+5.30\,T^{\frac{3}{2}}e^{\frac{-8.52}{T}}\big]\big[1-erf(2\sqrt{\alpha}\,T^{-\frac{1}{2}})\big]^2\Big)\notag \\
&+\dfrac{\dfrac{2.34}{T}\,e^{\frac{-2.34}{T}}+e^{\frac{-8.52}{T}}\big(2.65\,T^{\frac{1}{2}}+45.15\,T^{-\frac{1}{2}}\big)}{e^{\frac{-2.34}{T}}+5.30\, T^{\frac{1}{2}}e^{\frac{-8.52}{T} }}\notag \\
&+\dfrac{4\sqrt{\frac{\alpha}{\pi}}\,T^{-\frac{1}{2}}\,e^{-\frac{4\alpha}{T}}}{1-erf(2\sqrt{\alpha}\,T^{-\frac{1}{2}})}-\ln N\Big]\,.
\end{align}
Then entropy of a three-dimensional IR-deformed quantum bouncer is attained as
\begin{align}
S(N=1,T)=&2+\ln\big(\frac{\pi}{16\alpha}\big)+\ln\Big(\big[Te^{\frac{-2.34}{T}}+5.30\,T^{\frac{3}{2}}e^{\frac{-8.52}{T}}\big]\big[1-erf(2\sqrt{\alpha}\,T^{-\frac{1}{2}})\big]^2\Big)\notag \\
&+\dfrac{\dfrac{2.34}{T}\,e^{\frac{-2.34}{T}}+e^{\frac{-8.52}{T}}\big(2.65\,T^{\frac{1}{2}}+45.15\,T^{-\frac{1}{2}}\big)}{e^{\frac{-2.34}{T}}+5.30\, T^{\frac{1}{2}}e^{\frac{-8.52}{T} }}\notag \\
&+\dfrac{4\sqrt{\frac{\alpha}{\pi}}\,T^{-\frac{1}{2}}\,e^{-\frac{4\alpha}{T}}}{1-erf(2\sqrt{\alpha}\,T^{-\frac{1}{2}})}\,.
\end{align}
Considering the fact that $\alpha$\,(as a parameter that marks quantum gravity effects) has essentially a very small value due to the meaningful effect of IR modification at the large distances, the behavior of $S(T)$ for a three-dimensional IR-deformed quantum bouncer is depicted in FIG. 5. Focusing on this figure, a discrimination can be identified between the three-dimensional and one-dimensional cases. In the one-dimensional case, the particle freezes up almost in the temperature below $1$ K\,. However, this fact cannot be recognized in three-dimensional case, as the entropy in this case varies sharply below $1$ K\,. Such a different behavior may suggest the issue of thermodynamic dimensional reduction for a quantum bouncer in the three-dimensional case. To probe the claim more specifically, we attempt to calculate $U(N,T)$\,, the internal energy of a canonical ensemble of IR-deformed quantum bouncers, as follows
\begin{align}
\dfrac{U(N,T)}{N}=T-3\alpha+\dfrac{4\sqrt{\frac{\alpha}{\pi}}\,T^{\frac{1}{2}}\,e^{-\frac{4\alpha}{T}}}{1-erf\big(2\sqrt{\frac{\alpha}{T}}\big)}+\dfrac{2.34\,e^{\frac{-2.34}{T}}+e^{\frac{-8.52}{T}}\big(2.65\,T^{\frac{3}{2}}+45.15\,T^{\frac{1}{2}}\big)}{e^{\frac{-2.34}{T}}+5.30\, T^{\frac{1}{2}}\,e^{\frac{-8.52}{T} }}\,.
\end{align}
\begin{figure}[ht!]
\centering
\includegraphics[width=16cm]{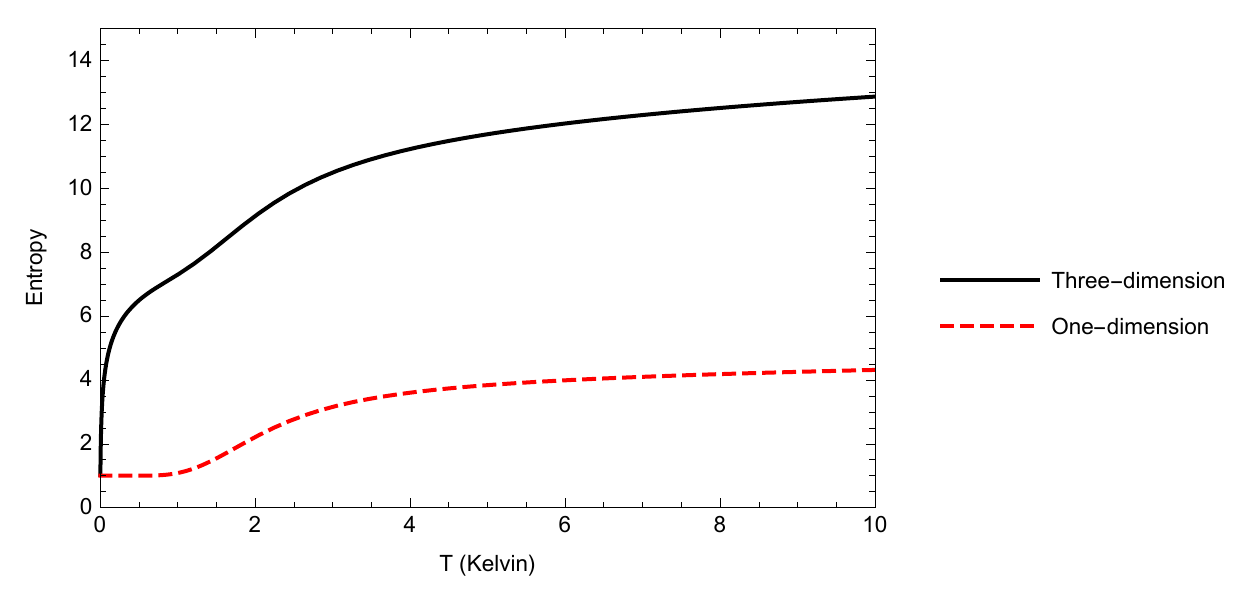}
\caption{\small {Entropy of a three-dimensional IR-deformed quantum bouncer (solid line) compared with a one-dimensional case (dashed line) setting $\alpha=0.0001$. As expected, the entropy and then information content of the physical system are increased by adding two extra degrees of freedom. Note the different behavior of the two curves below $1$ K\,.}}
\end{figure}
To move further, the asymptotic behavior of $\frac{U(N,T)}{N}$\, must be inspected as below
\begin{equation}\label{16}
\lim_{_{T \to 0}} \dfrac{U(N,T)}{N}=5\alpha+2.34\,,
\end{equation}
\begin{equation}\label{17}
\lim_{_{T \to \infty}} \dfrac{U(N,T)}{N}=\dfrac{3}{2}T+4\sqrt{\dfrac{\alpha T}{\pi}}\approx \dfrac{3}{2}T\,.
\end{equation}
The Eqs. (\ref{16}) and (\ref{17}) are interesting. Eq. (\ref{17}) shows that an IR-deformed quantum bouncer tends to behave classically as $T$\, becomes large, when IR modification fades away. More interestingly, Eq. (\ref{16}) implies two facts. Firstly, an IR modification leads to a minimal energy of a three-dimensional quantum bouncer which is more than that of one-dimensional case, i.e. $2.34-\alpha$, by $6\alpha$\,, for the impact of two new degrees of freedom. Secondly, this relation excitingly implies that a three-dimensional quantum bouncer experiences a thermodynamic dimensional reduction from $D=3$ to $D=2$ as considering IR regime of energy, in agreement with our claim. To be more specific, the mean internal energy of a one-dimensional IR-deformed quantum bouncer becomes independent of $T$\,below $1$\,K as shown in FIG. 4. This fact implies that the frozen dimension is $z$\,, the direction in which gravity affects. This can lead to the conclusion that gravity does not considerably affect the motion of a three-dimensional quantum bouncer within IR regime of energy. Then, the motion of a three-dimensional quantum bouncer becomes planar in $xy$\, plane within IR regime. This thermodynamic dimensional reduction can be explicitly identified in FIG. 7.
\begin{figure}[ht!]
\centering
\includegraphics[width=16cm]{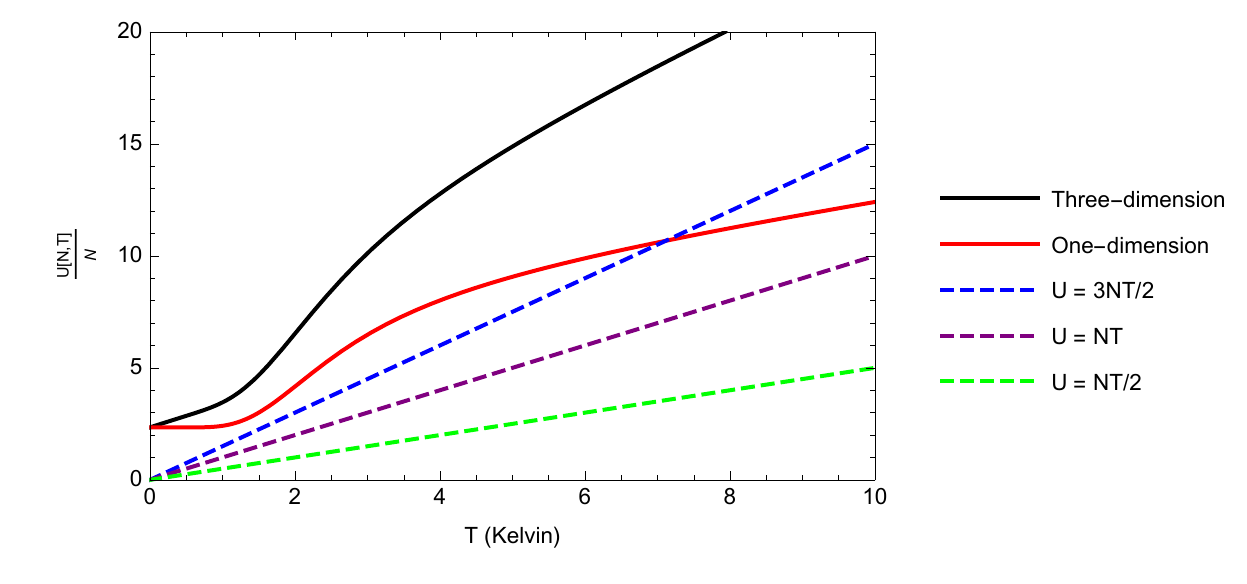}
\includegraphics[width=16cm]{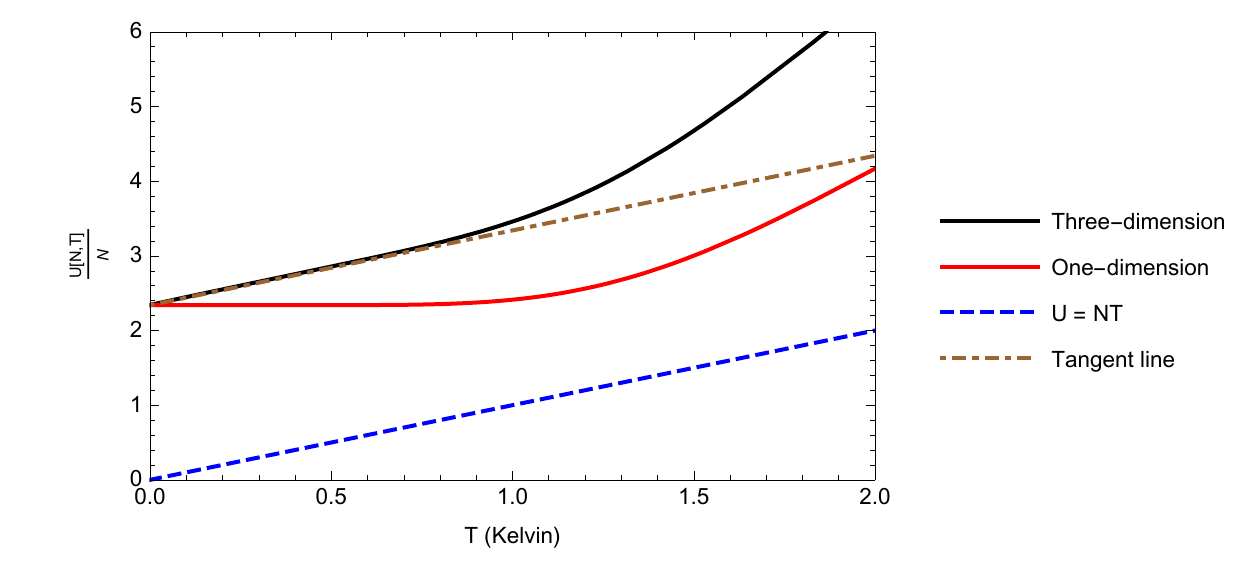}
\caption{\small {The mean internal energy of a three-dimensional IR-deformed quantum bouncer (black line) as compared with one-dimensional case (red line) setting $\alpha=0.0001$. A trace of thermodynamic dimensional reduction from $D=3$ to $D=2$ can be identified under $1$\,K as the line of $U=NT$\, becomes parallel to the tangent line for the three-dimensional case within this range of temperature.}}
\end{figure}
To more clarify the behavior of a three-dimensional quantum bouncer within IR regime, heat capacity can also be calculated as $C_{_{V}}=\frac{\partial\,U(N,T)}{\partial T}$\,.
\begin{align}
\dfrac{C_{_{V}}}{N}=1&+\sqrt{\dfrac{\alpha}{\pi}}\,\dfrac{e^{-\frac{4\alpha}{T}}\big(2\,T^{-\frac{1}{2}}+16\alpha\,T^{-\frac{3}{2}}\big)}{1-erf(2\sqrt{\alpha}\,T^{-\frac{1}{2}})}+\dfrac{8\alpha}{\pi}\,\dfrac{T^{-1}\,e^{-\frac{8\alpha}{T}}}{\big(1-erf(2\sqrt{\alpha}\,T^{-\frac{1}{2}})\big)^{2}}\notag \\
&+\dfrac{14.02\,T\,e^{-\frac{17.02}{T}}+e^{-\frac{8.52}{T}}\,(45.15\,T^{-\frac{1}{2}}+384.67\,T^{-\frac{3}{2}}+
3.97\,T^{\frac{1}{2}})}{\big(e^{-\frac{2.34}{T}}+5.30\,T^{\frac{1}{2}}\,e^{-\frac{8.52}{T}}\big)^{2}}\,.
\end{align}
The behavior of $C_{_{V}}(N,T)$\, is depicted in FIG. 7, suggesting again the thermodynamic dimensional reduction in IR regime that leads to a planar motion of a three-dimensional quantum bouncer.
\begin{figure}[ht!]
\centering
\includegraphics[width=16cm]{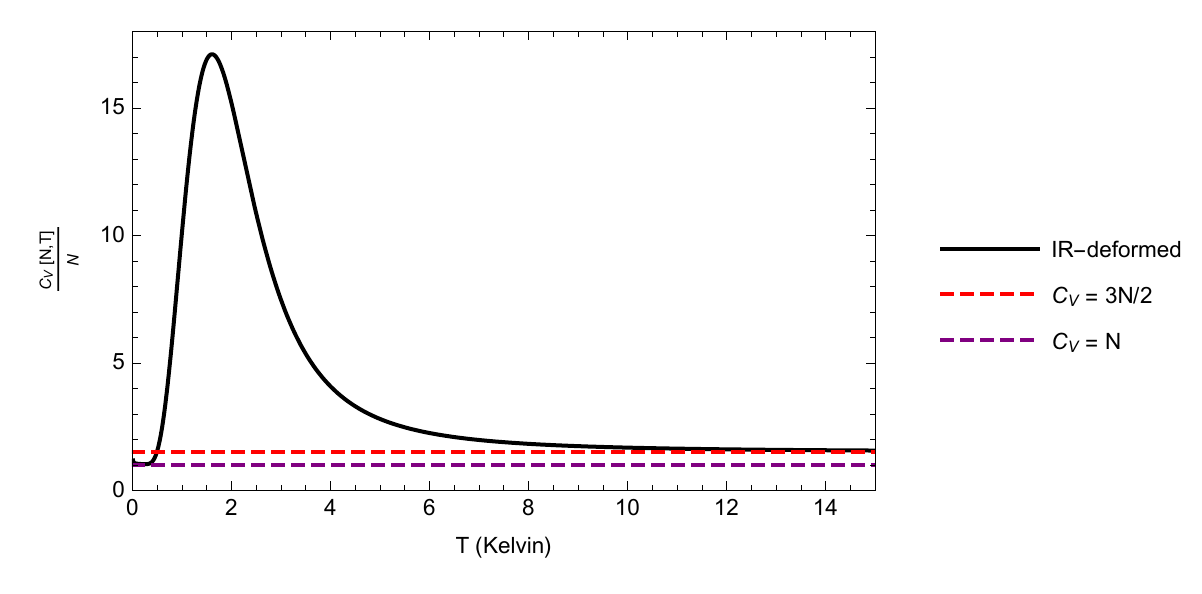}
\caption{\small {The mean heat capacity of a three-dimensional IR-deformed quantum bouncer for the case $\alpha=0.0001$. A trace of thermodynamic dimensional reduction is recognized under $1$\,K to two degrees of freedom. As temperature rises, a quantum bouncer tends to behave classically in three dimensions, since $C_{_{V}}$\, tends to $\dfrac{3}{2}\,N$, then IR modification fades away by rise of the temperature as expected. }}
\end{figure}
\section{Summary and Conclusion}
Ultracold neutrons have provided a promising probe to realize semi-classical quantum gravitational effects in low energies. These low energy neutrons can be theoretically modeled as quantum bouncers. While the high energy, ultraviolet regime of such a system has been studied within a UV-deformed quantum mechanics in Ref.~\cite{Pedram2011}, where the authors obtained some severe constraints on the quantum gravity parameter of the corresponding generalized uncertainty principle, the low energy, infrared regime of such a system has not been treated carefully so far. The present paper fills this gap. Low energy, infrared modified quantum mechanics is obviously relevant to the issue of ultracold neutrons. However, the superb fact is that this issue can be achieved via a phenomenological inspection of quantum gravitational effects in a low energy regime. Naturally, it is expectable to probe quantum gravitational effects in high energies, which belongs to the UV sector of quantum field theory. But, it has been shown that the low energy, infrared effects of semi-classical gravity, as a quantum gravitational manifestation, can be realized by an extended uncertainty principle that encodes a minimal measurable momentum~\cite{Mignemi2010}. This is relevant to the issue of ultracold neutrons and even seems necessary to be included in the famous Nesvizhevsky \emph{et al}. experiment~\cite{Nevi2002}. This is the main scenario in this paper. We considered the effects of an infrared cutoff, as a minimal measurable momentum, in the energy spectrum and eigenstates of a system of quantum bouncers, which theoretically models ultracold neutrons. The results are interesting and promising. Using an infrared modified Hamiltonian encoding a minimal measurable momentum, we derived energy spectrum and bound states of a one-dimensional quantum bouncer that models an ultracold neutron in one dimension. We used the first-order WKB approximation that is viable at this low energy limit. The calculated energy levels in this effective infrared regime are in agreement with the observed energy levels in the Nesvizhevsky \emph{et al.} experiment, with a negative constant deformation term as dependent on the deformation parameter. Our inspection of bound states suggests more localization of the deformed quantum bouncer as the deformation parameter increases, added by a slight reduction of the spatial frequency of the resulting deformed wave functions. Going forward, thermodynamics of a system of one-dimensional quantum bouncers is studied in order to make a comparison with the three-dimensional case. This comparison has revealed an interesting feature of this infrared problem: while the issue of dimensional reduction seems to be a high energy phenomenon in the realm of quantum gravity, we observed an effective thermodynamic dimensional reduction for such a low energy thermodynamic system. We showed that a system of three-dimensional quantum bouncers effectively acts as a two-dimensional system, realizing an effective dimensional reduction from $D=3$ to $D=2$. This is in contrast with the high energy regime dimensional reduction that admits reduction from $D=3$ to $D=1$. The importance of this achievement is that this is the first occasion for the observation of dimensional reduction in IR regime of a phenomenological quantum gravitational setup. \\


\begin{thebibliography}{100}
\bibitem{Hossenfelder2013}
S. Hossenfelder, Minimal Length Scale Scenarios for Quantum Gravity, Living Rev. Relativity \textbf{16}, (2013) 2.

\bibitem{Aguilar2013}
P. Aguilar, Y. Bonder and D. Sudarsky, Experimental search for a Lorentz invariant spacetime granularity: Possibilities and bounds,  Phys. Rev. D
\textbf{87}, (2013) 064007.

\bibitem{Scardigli1999}
F. Scardigli, Generalized Uncertainty Principle in Quantum Gravity from Micro-Black Hole Gedanken Experiment, Phys. Lett. B \textbf{452} (1999) 39.

\bibitem{Nozari2017}
M. Khodadi, K. Nozari and A. Hajizadeh, Some Astrophysical Aspects of a Schwarzschild Geometry Equipped with a Minimal Measurable Length, Phys. Lett. B \textbf{770}, (2017) 556.

\bibitem{Khodadi2018}
M. Khodadi, K. Nozari, A. Bhat and S. Mohsenian, Probing Planck Scale Spacetime By Cavity Opto-Atomic $^{87}$Rb Interferometry, [arXiv:1804.06389], To appear in Progress of Theoretical and Experimental Physics.

\bibitem{Hinrichsen1996}
H. Hinrichsen and A. Kempf, \textit{J. Math. Phys.} \textbf{37}, 2121 (1996).

\bibitem{Mignemi2010}
S. Mignemi, Extended uncertainty principle and the geometry of (anti)-de Sitter space, Mod. Phys. Lett. A \textbf{25} (2010) 1697.

\bibitem{Smolin2004}
J. Kowalski-Glikman and L. Smolin, Triply Special Relativity, Phys. Rev. D \textbf{70} (2004) 065020.

\bibitem{Mirza2009}
B. Mirza and M. Zarei, Minimal Uncertainty in Momentum: The Effects of IR Gravity on Quantum
Mechanics, Phys. Rev. D \textbf{79} (2009) 125007.

\bibitem{Zho2009}
T. Zhu, J. -R. Ren and M. -F. Li, Influence of Generalized and Extended Uncertainty Principle on Thermodynamics of
Friedmann-Robertson-Walker universe, Phys. Lett. B \textbf{674} (2009) 204.

\bibitem{Filho2016}
R. N. Costa Filho, J. P. M. Braga, J. H. S. Lira and J. S. Andrade Jr., Extended uncertainty from first principles, Phys. Lett. B \textbf{755} (2016) 367.

\bibitem{Mureika2019}
J. R. Mureika, Extended Uncertainty Principle black holes, Phys. Lett. B \textbf{789} (2019) 88.

\bibitem{Nozari2019}
K. Nozari and P. Dehghani, The role of an invariant IR cutoff in late time cosmological dynamics, Phys. Lett. B \textbf{792} (2019) 101.

\bibitem{Nozari2011} K. Nozari and P. Pedram, Minimal Length and Bouncing Particle Spectrum, Europhys. Lett. \textbf{92} (2010) 50013.

\bibitem{Pedram2011}
P. Pedram, K. Nozari and S. H. Taheri, The effects of minimal length and maximal momentum on the transition rate of ultra cold neutrons in gravitational field, JHEP \textbf{1103} (2011) 093.

\bibitem{Nevi2002}
V. V. Nesvizhevsky, H. G. B\"{o}rner, A. K. Petukhov, H. Abele, S. Bae{\ss}ler, F. J. Rue{\ss}, T. St\"{o}ferle, A. Westphal, A. M. Gagarski, G. A. Petrov and A. V. Strelkov, Quantum states of neutrons in the Earth's gravitational field, Nature \textbf{415} (2002) 297.

\bibitem{Nozari2015}
K. Nozari, V. Hosseinzadeh and M. A. Gorji, High temperature dimensional reduction in Snyder space, Phys. Lett. B \textbf{750} (2015) 218.

\bibitem{Carlip2017}
S. Carlip, Dimension and Dimensional Reduction in Quantum Gravity, Class. Quantum Grav. \textbf{34} (2017) 193001 [arXiv:1705.05417]. See also S. Carlip, Universe \textbf{5} (2019) 83.

\bibitem{valee2004}
O. Valee and M. Soares, Airy Functions and Applications to Physics, Imperial College Press (2004), London, UK.

\bibitem{Abramowitz1965}
M. Abramowitz and I. A. Stegun, Handbook of Mathematical Functions: with Formulas, Graphs, and Mathematical Tables, Dover Publications (1965), Washington, USA.

\end{thebibliography}
\end{document}